\documentclass[11pt,twocolumns,article,nofontune]{IEEEtran}
\usepackage{fancyhdr} 
\usepackage[USenglish,american]{babel}
\usepackage{epsfig,graphics,graphicx,latexsym,amsmath,amssymb}
\usepackage{multirow}
\usepackage[usenames,dvipsnames]{color}
\usepackage{xcolor}
\newcommand{\rev}[1]{\textcolor{black}{#1}}
\usepackage[T1]{fontenc}
\usepackage{url}
\usepackage{array}
\usepackage{booktabs,siunitx}
\usepackage{makecell}
\usepackage[small]{caption}

\usepackage{pgfplots}
\pgfplotsset{compat=1.16}
\usepgfplotslibrary{units}
\usepackage[acronyms,nonumberlist,nopostdot,nomain,nogroupskip,acronymlists={hidden}]{glossaries}
\newglossary[algh]{hidden}{acrh}{acnh}{Hidden Acronyms}
\glsdisablehyper
\newacronym{3gpp}{3GPP}{3rd Generation Partnership Project}
\newacronym{4g}{4G}{4th generation}
\newacronym{5g}{5G}{5th generation}
\newacronym{6g}{6G}{6th generation}
\newacronym{5gc}{5GC}{5G Core}
\newacronym{adc}{ADC}{Analog to Digital Converter}
\newacronym{aerpaw}{AERPAW}{Aerial Experimentation and Research Platform for Advanced Wireless}
\newacronym{ai}{AI}{Artificial Intelligence}
\newacronym{aimd}{AIMD}{Additive Increase Multiplicative Decrease}
\newacronym{am}{AM}{Acknowledged Mode}
\newacronym{amc}{AMC}{Adaptive Modulation and Coding}
\newacronym{amf}{AMF}{Access and Mobility Management Function}
\newacronym{aoa}{AoA}{Angle of Arrival}
\newacronym{aops}{AOPS}{Adaptive Order Prediction Scheduling}
\newacronym{api}{API}{Application Programming Interface}
\newacronym{apn}{APN}{Access Point Name}
\newacronym{aqm}{AQM}{Active Queue Management}
\newacronym{ausf}{AUSF}{Authentication Server Function}
\newacronym{avc}{AVC}{Advanced Video Coding}
\newacronym{awgn}{AGWN}{Additive White Gaussian Noise}
\newacronym{balia}{BALIA}{Balanced Link Adaptation Algorithm}
\newacronym{bbu}{BBU}{Base Band Unit}
\newacronym{bdp}{BDP}{Bandwidth-Delay Product}
\newacronym{ber}{BER}{Bit Error Rate}
\newacronym{bf}{BF}{Beamforming}
\newacronym{bler}{BLER}{Block Error Rate}
\newacronym{brr}{BRR}{Bayesian Ridge Regressor}
\newacronym{bsr}{BSR}{Buffer Status Report}
\newacronym{bs}{BS}{Base Station}
\newacronym{bss}{BSS}{Business Support System}
\newacronym{ca}{CA}{Carrier Aggregation}
\newacronym{caas}{CaaS}{Connectivity-as-a-Service}
\newacronym{cb}{CB}{Code Block}
\newacronym{cc}{CC}{Congestion Control}
\newacronym{ccid}{CCID}{Congestion Control ID}
\newacronym{cco}{CC}{Carrier Component}
\newacronym{cdd}{CDD}{Cyclic Delay Diversity}
\newacronym{cdf}{CDF}{Cumulative Distribution Function}
\newacronym{cdn}{CDN}{Content Distribution Network}
\newacronym{csi-rs}{CSI-RS}{Channel State Information Reference Signal}
\newacronym{cir}{CIR}{Channel Impulse Response}
\newacronym{cn}{CN}{Core Network}
\newacronym{codel}{CoDel}{Controlled Delay Management}
\newacronym{comac}{COMAC}{Converged Multi-Access and Core}
\newacronym{cord}{CORD}{Central Office Re-architected as a Datacenter}
\newacronym{cornet}{CORNET}{COgnitive Radio NETwork}
\newacronym{cosmos}{COSMOS}{Cloud Enhanced Open Software Defined Mobile Wireless Testbed for City-Scale Deployment}
\newacronym{cots}{COTS}{Commercial Off-the-Shelf}
\newacronym{cp}{CP}{Cyclic Prefix}
\newacronym{cpu}{CPU}{Central Processing Unit}
\newacronym{cqi}{CQI}{Channel Quality Information}
\newacronym{cr}{CR}{Cognitive Radio}
\newacronym{cran}{CRAN}{Cloud \gls{ran}}
\newacronym{crc}{CRC}{Cyclic Redundancy Check}
\newacronym{crlb}{CRLB}{Cramér–Rao Lower Bound}
\newacronym{crs}{CRS}{Cell Reference Signal}
\newacronym{csi}{CSI}{Channel State Information}
\newacronym{csirs}{CSI-RS}{Channel State Information - Reference Signal}
\newacronym{cu}{CU}{Central Unit}
\newacronym{d2tcp}{D$^2$TCP}{Deadline-aware Data center TCP}
\newacronym{d3}{D$^3$}{Deadline-Driven Delivery}
\newacronym{dac}{DAC}{Digital to Analog Converter}
\newacronym{dag}{DAG}{Directed Acyclic Graph}
\newacronym{darpa}{DARPA}{Defense Advanced Research Projects Agency}
\newacronym{das}{DAS}{Distributed Antenna System}
\newacronym{dash}{DASH}{Dynamic Adaptive Streaming over HTTP}
\newacronym{dc}{DC}{Dual Connectivity}
\newacronym{dccp}{DCCP}{Datagram Congestion Control Protocol}
\newacronym{dce}{DCE}{Direct Code Execution}
\newacronym{dci}{DCI}{Downlink Control Information}
\newacronym{dcl}{DCL}{Dear Colleague Letter}
\newacronym{dctcp}{DCTCP}{Data Center TCP}
\newacronym{dl}{DL}{Downlink}
\newacronym{dmr}{DMR}{Deadline Miss Ratio}
\newacronym{dmrs}{DMRS}{DeModulation Reference Signal}
\newacronym{drlcc}{DRL-CC}{Deep Reinforcement Learning Congestion Control}
\newacronym{drs}{DRS}{Discovery Reference Signal}
\newacronym{du}{DU}{Distributed Unit}
\newacronym{e2e}{E2E}{end-to-end}
\newacronym{ecaas}{ECaaS}{Edge-Cloud-as-a-Service}
\newacronym{ecn}{ECN}{Explicit Congestion Notification}
\newacronym{edf}{EDF}{Earliest Deadline First}
\newacronym{embb}{eMBB}{Enhanced Mobile Broadband}
\newacronym{empower}{EMPOWER}{EMpowering transatlantic PlatfOrms for advanced WirEless Research}
\newacronym{enb}{eNB}{evolved Node Base}
\newacronym{endc}{EN-DC}{E-UTRAN-\gls{nr} \gls{dc}}
\newacronym{epc}{EPC}{Evolved Packet Core}
\newacronym{eps}{EPS}{Evolved Packet System}
\newacronym{es}{ES}{Edge Server}
\newacronym{etsi}{ETSI}{European Telecommunications Standards Institute}
\newacronym[firstplural=Estimated Times of Arrival (ETAs)]{eta}{ETA}{Estimated Time of Arrival}
\newacronym{eutran}{E-UTRAN}{Evolved Universal Terrestrial Access Network}
\newacronym{faas}{FaaS}{Function-as-a-Service}
\newacronym{fapi}{FAPI}{Functional Application Platform Interface}
\newacronym{fcc}{FCC}{Federal Communications Commission}
\newacronym{fdd}{FDD}{Frequency Division Duplexing}
\newacronym{fdm}{FDM}{Frequency Division Multiplexing}
\newacronym{fdma}{FDMA}{Frequency Division Multiple Access}
\newacronym{fed4fire}{FED4FIRE+}{Federation 4 Future Internet Research and Experimentation Plus}
\newacronym{fir}{FIR}{Finite Impulse Response}
\newacronym{fit}{FIT}{Future \acrlong{iot}}
\newacronym{fmcw}{FMCW}{Frequency-Modulated Continuous-Wave}  
\newacronym{fft}{FFT}{Fourier Transform}
\newacronym{fpga}{FPGA}{Field Programmable Gate Array}
\newacronym{fr2}{FR2}{Frequency Range 2}
\newacronym{fs}{FS}{Fast Switching}
\newacronym{fscc}{FSCC}{Flow Sharing Congestion Control}
\newacronym{ftp}{FTP}{File Transfer Protocol}
\newacronym{fw}{FW}{Flow Window}
\newacronym{ge}{GE}{Gaussian Elimination}
\newacronym{gnb}{gNB}{Base Station}
\newacronym{gop}{GOP}{Group of Pictures}
\newacronym{gpr}{GPR}{Gaussian Process Regressor}
\newacronym{gps}{GPS}{Global Positioning System}
\newacronym{gpu}{GPU}{Graphics Processing Unit}
\newacronym{gtp}{GTP}{GPRS Tunneling Protocol}
\newacronym{gtpc}{GTP-C}{GPRS Tunnelling Protocol Control Plane}
\newacronym{gtpu}{GTP-U}{GPRS Tunnelling Protocol User Plane}
\newacronym{gtpv2c}{GTPv2-C}{\gls{gtp} v2 - Control}
\newacronym{gw}{GW}{Gateway}
\newacronym{harq}{HARQ}{Hybrid Automatic Repeat reQuest}
\newacronym{hetnet}{HetNet}{Heterogeneous Network}
\newacronym{hh}{HH}{Hard Handover}
\newacronym{hol}{HOL}{Head-of-Line}
\newacronym{hqf}{HQF}{Highest-quality-first}
\newacronym{hss}{HSS}{Home Subscription Server}
\newacronym{http}{HTTP}{HyperText Transfer Protocol}
\newacronym{ia}{IA}{Initial Access}
\newacronym{iab}{IAB}{Integrated Access and Backhaul}
\newacronym{ic}{IC}{Incident Command}
\newacronym{ietf}{IETF}{Internet Engineering Task Force}
\newacronym{imsi}{IMSI}{International Mobile Subscriber Identity}
\newacronym{imt}{IMT}{International Mobile Telecommunication}
\newacronym{iot}{IoT}{Internet of Things}
\newacronym{ip}{IP}{Internet Protocol}
\newacronym{isac}{ISAC}{Integrated Sensing and Communication}
\newacronym{isi}{ISI}{Intersymbol Interference}
\newacronym{itu}{ITU}{International Telecommunication Union}
\newacronym{kpi}{KPI}{Key Performance Indicator}
\newacronym{kvm}{KVM}{Kernel-based Virtual Machine}
\newacronym{ldpc}{LDPC}{Low-Density Parity Check}
\newacronym{los}{LoS}{Line-of-Sight}
\newacronym{lsm}{LSM}{Link-to-System Mapping}
\newacronym{lstm}{LSTM}{Long Short Term Memory}
\newacronym{lte}{LTE}{Long Term Evolution}
\newacronym{lxc}{LXC}{Linux Container}
\newacronym{m2m}{M2M}{Machine to Machine}
\newacronym{mac}{MAC}{Medium Access Control}
\newacronym{manet}{MANET}{Mobile Ad Hoc Network}
\newacronym{mano}{MANO}{Management and Orchestration}
\newacronym{mc}{MC}{Multi-Connectivity}
\newacronym{mcc}{MCC}{Mobile Cloud Computing}
\newacronym{mchem}{MCHEM}{Massive Channel Emulator}
\newacronym{mcs}{MCS}{Modulation and Coding Scheme}
\newacronym{mcss}{MCS}{Modulation and Coding Schemes}
\newacronym{mec}{MEC}{Multi-access Edge Computing}
\newacronym{mec2}{MEC}{Mobile Edge Cloud}
\newacronym{mfc}{MFC}{Mobile Fog Computing}
\newacronym{mi}{MI}{Mutual Information}
\newacronym{mib}{MIB}{Master Information Block}
\newacronym{miesm}{MIESM}{Mutual Information Based Effective SINR}
\newacronym{mimo}{MIMO}{Multiple Input, Multiple Output}
\newacronym{mgen}{MGEN}{Multi-Generator}
\newacronym{ml}{ML}{Machine Learning}
\newacronym{mlr}{MLR}{Maximum-local-rate}
\newacronym[plural=\gls{mme}s,firstplural=Mobility Management Entities (MMEs)]{mme}{MME}{Mobility Management Entity}
\newacronym{mse}{MSE}{Mean Square Error}
\newacronym{mmtc}{mMTC}{Massive Machine-Type Communications}
\newacronym{mmwave}{mmWave}{millimeter wave}
\newacronym{mpdccp}{MP-DCCP}{Multipath Datagram Congestion Control Protocol}
\newacronym{mptcp}{MPTCP}{Multipath TCP}
\newacronym{mr}{MR}{Maximum Rate}
\newacronym{mrdc}{MR-DC}{Multi \gls{rat} \gls{dc}}
\newacronym{mss}{MSS}{Maximum Segment Size}
\newacronym{mt}{MT}{Mobile Termination}
\newacronym{mtd}{MTD}{Machine-Type Device}
\newacronym{mtu}{MTU}{Maximum Transmission Unit}
\newacronym{mumimo}{MU-MIMO}{Multi-user \gls{mimo}}
\newacronym{mvno}{MVNO}{Mobile Virtual Network Operator}
\newacronym{nalu}{NALU}{Network Abstraction Layer Unit}
\newacronym{nas}{NAS}{Network Attached Storage}
\newacronym{nbiot}{NB-IoT}{Narrow Band IoT}
\newacronym{nfv}{NFV}{Network Function Virtualization}
\newacronym{nfvi}{NFVI}{Network Function Virtualization Infrastructure}
\newacronym{nic}{NIC}{Network Interface Card}
\newacronym{nlos}{NLOS}{Non-Line-of-Sight}
\newacronym{now}{NOW}{Non Overlapping Window}
\newacronym{nrdz}{NRDZ}{National Radio Dynamic Zone}
\newacronym{nsf}{NSF}{National Science Foundation}
\newacronym{nsm}{NSM}{Network Service Mesh}
\newacronym[type=hidden]{nr}{NR}{New Radio}
\newacronym{nrf}{NRF}{Network Repository Function}
\newacronym{nsa}{NSA}{Non Stand Alone}
\newacronym{nse}{NSE}{Network Slicing Engine}
\newacronym{nssf}{NSSF}{Network Slice Selection Function}
\newacronym{ntp}{NTP}{Network Time Protocol}
\newacronym{o2i}{O2I}{Outdoor to Indoor}
\newacronym{oai}{OAI}{OpenAirInterface}
\newacronym{oaicn}{OAI-CN}{\gls{oai} \acrlong{cn}}
\newacronym{oairan}{OAI-RAN}{\acrlong{oai} \acrlong{ran}}
\newacronym{oam}{OAM}{Operations, Administration and Maintenance}
\newacronym{ofdm}{OFDM}{Orthogonal Frequency Division Multiplexing}
\newacronym{olia}{OLIA}{Opportunistic Linked Increase Algorithm}
\newacronym{omec}{OMEC}{Open Mobile Evolved Core}
\newacronym{onap}{ONAP}{Open Network Automation Platform}
\newacronym{onf}{ONF}{Open Networking Foundation}
\newacronym{onos}{ONOS}{Open Networking Operating System}
\newacronym{oom}{OOM}{\gls{onap} Operations Manager}
\newacronym{opnfv}{OPNFV}{Open Platform for \gls{nfv}}
\newacronym[type=hidden]{oran}{O-RAN}{Open Radio Access Network}
\newacronym{orbit}{ORBIT}{Open-Access Research Testbed for Next-Generation Wireless Networks}
\newacronym{os}{OS}{Operating System}
\newacronym{oss}{OSS}{Operations Support System}
\newacronym{otfs}{OTFS}{Orthogonal Time Frequency Space}
\newacronym{pa}{PA}{Position-aware}
\newacronym{pase}{PASE}{Prioritization, Arbitration, and Self-adjusting Endpoints}
\newacronym{pawr}{PAWR}{Platforms for Advanced Wireless Research}
\newacronym{pbch}{PBCH}{Physical Broadcast Channel}
\newacronym{pcef}{PCEF}{Policy and Charging Enforcement Function}
\newacronym{pcfich}{PCFICH}{Physical Control Format Indicator Channel}
\newacronym{pcrf}{PCRF}{Policy and Charging Rules Function}
\newacronym{pdcch}{PDCCH}{Physical Downlink Control Channel}
\newacronym{pdcp}{PDCP}{Packet Data Convergence Protocol}
\newacronym{pdsch}{PDSCH}{Physical Downlink Shared Channel}
\newacronym{pdu}{PDU}{Packet Data Unit}
\newacronym{pf}{PF}{Proportional Fair}
\newacronym{pgw}{PGW}{Packet Gateway}
\newacronym{phich}{PHICH}{Physical Hybrid ARQ Indicator Channel}
\newacronym{phy}{PHY}{Physical}
\newacronym{pmch}{PMCH}{Physical Multicast Channel}
\newacronym{pmi}{PMI}{Precoding Matrix Indicators}
\newacronym{powder}{POWDER}{Platform for Open Wireless Data-driven Experimental Research}
\newacronym{ppo}{PPO}{Proximal Policy Optimization}
\newacronym{ppp}{PPP}{Poisson Point Process}
\newacronym{prach}{PRACH}{Physical Random Access Channel}
\newacronym{prb}{PRB}{Physical Resource Block}
\newacronym{prs}{PRS}{Positioning Reference Signal}
\newacronym{psnr}{PSNR}{Peak Signal to Noise Ratio}
\newacronym{pss}{PSS}{Primary Synchronization Signal}
\newacronym{pucch}{PUCCH}{Physical Uplink Control Channel}
\newacronym{pusch}{PUSCH}{Physical Uplink Shared Channel}
\newacronym{qam}{QAM}{Quadrature Amplitude Modulation}
\newacronym{qci}{QCI}{\gls{qos} Class Identifier}
\newacronym{qoe}{QoE}{Quality of Experience}
\newacronym{qos}{QoS}{Quality of Service}
\newacronym{quic}{QUIC}{Quick UDP Internet Connections}
\newacronym{ra}{RA}{Random Access}
\newacronym{rach}{RACH}{Random Access Channel}
\newacronym{ran}{RAN}{Radio Access Network}
\newacronym{rar}{RAR}{Random Access Response}
\newacronym[firstplural=Radio Access Technologies (RATs)]{rat}{RAT}{Radio Access Technology}
\newacronym{rcn}{RCN}{Research Coordination Network}
\newacronym{rec}{REC}{Radio Edge Cloud}
\newacronym{red}{RED}{Random Early Detection}
\newacronym{renew}{RENEW}{Reconfigurable Eco-system for Next-generation End-to-end Wireless}
\newacronym{re}{RE}{Resource Element}
\newacronym{rf}{RF}{Radio Frequency}
\newacronym{rfc}{RFC}{Request for Comments}
\newacronym{rfr}{RFR}{Random Forest Regressor}
\newacronym{ric}{RIC}{RAN Intelligent Controller}
\newacronym{rlc}{RLC}{Radio Link Control}
\newacronym{rlf}{RLF}{Radio Link Failure}
\newacronym{rlnc}{RLNC}{Random Linear Network Coding}
\newacronym{rmse}{RMSE}{Root Mean Squared Error}
\newacronym{rnis}{RNIS}{Radio Network Information Service}
\newacronym{rr}{RR}{Round Robin}
\newacronym{rrc}{RRC}{Radio Resource Control}
\newacronym{rrm}{RRM}{Radio Resource Management}
\newacronym{rru}{RRU}{Remote Radio Unit}
\newacronym{rs}{RS}{Remote Server}
\newacronym{rsrp}{RSRP}{Reference Signal Received Power}
\newacronym{rsrq}{RSRQ}{Reference Signal Received Quality}
\newacronym{rss}{RSS}{Received Signal Strength}
\newacronym{rssi}{RSSI}{Received Signal Strength Indicator}
\newacronym{rtt}{RTT}{Round Trip Time}
\newacronym{ru}{RU}{Radio Unit}
\newacronym{rw}{RW}{Receive Window}
\newacronym{rx}{RX}{Receiver}
\newacronym{s1ap}{S1AP}{S1 Application Protocol}
\newacronym{sa}{SA}{standalone}
\newacronym{sack}{SACK}{Selective Acknowledgment}
\newacronym{sap}{SAP}{Service Access Point}
\newacronym{sc2}{SC2}{Spectrum Collaboration Challenge}
\newacronym{scef}{SCEF}{Service Capability Exposure Function}
\newacronym{sch}{SCH}{Secondary Cell Handover}
\newacronym{scoot}{SCOOT}{Split Cycle Offset Optimization Technique}
\newacronym{sctp}{SCTP}{Stream Control Transmission Protocol}
\newacronym{sdap}{SDAP}{Service Data Adaptation Protocol}
\newacronym{sdk}{SDK}{Software Development Kit}
\newacronym{sdm}{SDM}{Space Division Multiplexing}
\newacronym{sdma}{SDMA}{Spatial Division Multiple Access}
\newacronym{sdn}{SDN}{Software-defined Networking}
\newacronym{sdr}{SDR}{Software-defined Radio}
\newacronym{seba}{SEBA}{SDN-Enabled Broadband Access}
\newacronym{sgsn}{SGSN}{Serving GPRS Support Node}
\newacronym{sgw}{SGW}{Service Gateway}
\newacronym{si}{SI}{Study Item}
\newacronym{sib}{SIB}{Secondary Information Block}
\newacronym{sic}{SIC}{Successive Interference Cancellation}
\newacronym{sinr}{SINR}{Signal to Interference plus Noise Ratio}
\newacronym{sip}{SIP}{Session Initiation Protocol}
\newacronym{siso}{SISO}{Single Input, Single Output}
\newacronym{sla}{SLA}{Service Level Agreement}
\newacronym{sm}{SM}{Saturation Mode}
\newacronym{smf}{SMF}{Session Management Function}
\newacronym{smo}{SMO}{Service Management and Orchestration}
\newacronym{sms}{SMS}{Short Message Service}
\newacronym{smsgmsc}{SMS-GMSC}{\gls{sms}-Gateway}
\newacronym{snr}{SNR}{Signal-to-Noise-Ratio}
\newacronym{son}{SON}{Self-Organizing Network}
\newacronym{sptcp}{SPTCP}{Single Path TCP}
\newacronym{srb}{SRB}{Service Radio Bearer}
\newacronym{srn}{SRN}{Standard Radio Node}
\newacronym{srs}{SRS}{Sounding Reference Signal}
\newacronym{ss}{SS}{Synchronization Signal}
\newacronym{ssb}{SSB}{Synchronization Signal Block}
\newacronym{sss}{SSS}{Secondary Synchronization Signal}
\newacronym{st}{ST}{Spanning Tree}
\newacronym{svc}{SVC}{Scalable Video Coding}
\newacronym{ta}{TA}{Timing Advance}
\newacronym{tb}{TB}{Transport Block}
\newacronym{tcp}{TCP}{Transmission Control Protocol}
\newacronym{tdd}{TDD}{Time Division Duplexing}
\newacronym{tdm}{TDM}{Time Division Multiplexing}
\newacronym{tdma}{TDMA}{Time Division Multiple Access}
\newacronym{tfl}{TfL}{Transport for London}
\newacronym{tfrc}{TFRC}{TCP-Friendly Rate Control}
\newacronym{tft}{TFT}{Traffic Flow Template}
\newacronym{tgen}{TGEN}{Traffic Generator}
\newacronym{tip}{TIP}{Telecom Infra Project}
\newacronym{tm}{TM}{Transparent Mode}
\newacronym{to}{TO}{Telco Operator}
\newacronym{tr}{TR}{Technical Report}
\newacronym{trp}{TRP}{Transmitter Receiver Pair}
\newacronym{ts}{TS}{Technical Specification}
\newacronym{tti}{TTI}{Transmission Time Interval}
\newacronym{ttt}{TTT}{Time-to-Trigger}
\newacronym{tx}{TX}{Transmitter}
\newacronym{uas}{UAS}{Unmanned Aerial System}
\newacronym{uav}{UAV}{Unmanned Aerial Vehicle}
\newacronym{udm}{UDM}{Unified Data Management}
\newacronym{udp}{UDP}{User Datagram Protocol}
\newacronym{udr}{UDR}{Unified Data Repository}
\newacronym{ue}{UE}{User Equipment}
\newacronym{uhd}{UHD}{\gls{usrp} Hardware Driver}
\newacronym{ul}{UL}{Uplink}
\newacronym{ultdoa}{UL-TDoA}{Uplink Time Difference of Arrival}
\newacronym{um}{UM}{Unacknowledged Mode}
\newacronym{uml}{UML}{Unified Modeling Language}
\newacronym{upa}{UPA}{Uniform Planar Array}
\newacronym{upf}{UPF}{User Plane Function}
\newacronym{urllc}{URLLC}{Ultra Reliable and Low Latency Communication}
\newacronym{usa}{U.S.}{United States}
\newacronym{usim}{USIM}{Universal Subscriber Identity Module}
\newacronym{usrp}{USRP}{Universal Software Radio Peripheral}
\newacronym{utc}{UTC}{Urban Traffic Control}
\newacronym{vim}{VIM}{Virtualization Infrastructure Manager}
\newacronym{vm}{VM}{Virtual Machine}
\newacronym{vnf}{VNF}{Virtual Network Function}
\newacronym{volte}{VoLTE}{Voice over \gls{lte}}
\newacronym{voltha}{VOLTHA}{Virtual OLT HArdware Abstraction}
\newacronym{vr}{VR}{Virtual Reality}
\newacronym{vran}{vRAN}{Virtualized \gls{ran}}
\newacronym{vss}{VSS}{Video Streaming Server}
\newacronym{wbf}{WBF}{Wired Bias Function}
\newacronym{wf}{WF}{Wired-first}
\newacronym{wlan}{WLAN}{Wireless Local Area Network}
\newacronym{osm}{OSM}{Open Source \gls{nfv} Management and Orchestration}
\newacronym{pnf}{PNF}{Physical Network Function}
\newacronym{drl}{DRL}{Deep Reinforcement Learning}
\newacronym{mtc}{MTC}{Machine-type Communications}

\newacronym{cif}{CI}{cyberinfrastructure}
\newacronym{sonic}{SONiC}{Software for Open Networking in the Cloud}
\newacronym{ocp}{OCP}{Open Compute Project}
\newacronym{snmp}{SNMP}{Simple Network Management Protocol}
\newacronym{raid}{RAID}{redundant array of independent disks}
\newacronym{nfs}{NFS}{Network File Storage}
\newacronym{ci}{CI}{Continuous Integration}
\newacronym{cd}{CD}{Continuous Deployment}
\newacronym{dtn}{DTN}{Data Transfer Node}

\newacronym{dns}{DNS}{Domain Name Service}
\newacronym{nrpe}{NRPE}{Nagios Remote Plugin Executor}
\newacronym{ldap}{LDAP}{Lightweight Directory Access Protocol}
\newacronym{lan}{LAN}{Local Area Network}
\newacronym{vlan}{VLAN}{Virtual LAN}

\newacronym{ipmi}{IPMI}{Intelligent Platform Management Interface}
\newacronym{tor}{ToR}{Top-of-the-Rack}
\newacronym{lmn}{LMN}{Large Memory Node}
\newacronym{bgp}{BGP}{Border Gateway Protocol}
\newacronym{dhcp}{DHCP}{Dynamic Host Configuration Protocol}
\newacronym{vrf}{VRF}{Virtual Routing and Forwarding}
\newacronym{vpn}{VPN}{Virtual Private Network}
\newacronym{rma}{RMA}{Return Merchandise Authorization}
\newacronym{hpc}{HPC}{High Performance Compute}

\newacronym{nu}{NU}{Northeastern University}
\newacronym{asic}{ASIC}{Application-specific Integrated Circuit}
\newacronym{rdma}{RDMA}{Remote Direct Memory Access}
\newacronym{roce}{RoCE}{RDMA over Converged Ethernet}
\newacronym{ovs}{OVS}{Open vSwitch}
\newacronym{frr}{FRR}{Free Range Routing}
\newacronym{ups}{UPS}{Uninterruptible Power Supply}

\newacronym{ntia}{NTIA}{National Telecommunications and Information Administration}
\newacronym{pii}{PII}{Personal and Identifiable Information}
\newacronym{irb}{IRB}{Institutional Review Board}
\newacronym{doi}{DOI}{Digital Object Identifier}

\newacronym{sdo}{SDO}{Standard-Development Organization}
\newacronym{ose}{OSE}{Open Source Ecosystem}
\newacronym{osc}{OSC}{O-RAN Software Community}
\newacronym{dop}{DOP}{Director of Operations}
\newacronym{pm}{PM}{Program Manager}
\newacronym{excom}{EXCOM}{Executive Committee}
\newacronym{iiot}{IIoT}{Industrial \gls{iot}}
\newacronym{lf}{LF}{Linux Foundation}

\newacronym{wiot}{WIoT}{Institute for the Wireless Internet of Things}

\newacronym{otic}{OTIC}{Open Testing \& Integration Centre}

\newacronym{nofo}{NOFO}{Notice of Funding Opportunity}

\newacronym{onr}{ONR}{Office of Naval Research}
\newacronym{afosr}{AFOSR}{Air Force Office of Scientific Research}
\newacronym{afrl}{AFRL}{Air Force Research Laboratory}
\newacronym{arl}{ARL}{Army Research Laboratory}

\newacronym{arc}{ARC}{Aerial Research Cloud}

\newacronym{mno}{MNO}{Mobile Network Operator}
\newacronym{ct}{CT}{Continuous Testing}
\newacronym{oci}{OCI}{Open Container Initiative}
\newacronym{macsec}{MACsec}{Media Access Control Security}
\newacronym{pt}{PT}{Plain Text}
\newacronym{cuda}{CUDA}{Compute Unified Device Architecture}
\newacronym{cbrs}{CBRS}{Citizen Broadband Radio Service}
\newacronym{sas}{SAS}{Spectrum Access System}
\newacronym{rfi}{RFI}{Radio-Frequency Interference}
\newacronym{pal}{PAL}{Priority Access License}
\newacronym{gaa}{GAA}{General Authorized Access}
\newacronym{esc}{ESC}{Environmental Sensing Capability}
\newacronym{ota}{OTA}{Over-the-Air}
\newacronym{gdpr}{GDPR}{GDPR TODO}
\newacronym{ccpa}{CCPA}{CCPA TODO}
\newacronym{lcm}{LCM}{Life-Cycle Management}
\newacronym{ott}{OTT}{Over-the-Top}
\newacronym{ngrg}{NGRG}{Next Generation Research Group}
\newacronym{nf}{NF}{Network Function}
\newacronym{cnn}{CNN}{Convolutional Neural Network}
\newacronym{kpm}{KPM}{Key Performance Measurement}
\newacronym{rcs}{RCS}{Radar Cross-Section}
\title{Enabling Programmable Inference and ISAC\\at the 6GR Edge with dApps
\thanks{This paper is partially supported by the U.S. NSF under award CNS-2434081.}}

\author{\IEEEauthorblockN{Michele Polese, Rajeev Gangula, Tommaso Melodia\\
}
\IEEEauthorblockN{Institute for Intelligent Networked Systems, Northeastern University, Boston, MA, U.S.A.}
}

\begin{document}

\maketitle

\begin{abstract}
The convergence of communication, sensing, and \gls{ai} in the \gls{ran} offers compelling economic advantages through shared spectrum and infrastructure.
How can inference and sensing be integrated in the \gls{ran} infrastructure \emph{at a system level?}
Current abstractions in O-RAN and 3GPP lack the interfaces and capabilities to support (i) a dynamic life cycle for inference and \gls{isac} algorithms, whose requirements and sensing targets may change over time and across sites; (ii) pipelines for \gls{ai}-driven \gls{isac}, which need complex data flows, training, and testing; (iii) dynamic device and stack configuration to balance trade-offs between connectivity, sensing, and inference services.
This paper analyzes the role of a programmable, software-driven, open \gls{ran} in enabling the intelligent edge for 5G and 6G systems. We identify real-time user-plane data exposure, open interfaces for plug-and-play inference and \gls{isac} models, closed-loop control, and \gls{ai} pipelines as elements that evolutions of the O-RAN architecture can uniquely provide. Specifically, we describe how dApps---a real-time, user-plane extension of O-RAN---and a hierarchy of controllers enable real-time \gls{ai} inference and \gls{isac}. Experimental results on an open-source RAN testbed demonstrate the value of exposing I/Q samples and real-time RAN telemetry to dApps for sensing applications.
\end{abstract}

% \vspace{-.6cm}
%---------------------------------------------%
    \section{Introduction} \label{sec:intro}
%---------------------------------------------%

\begin{figure*}[t]
\centering
\includegraphics[width=\textwidth]{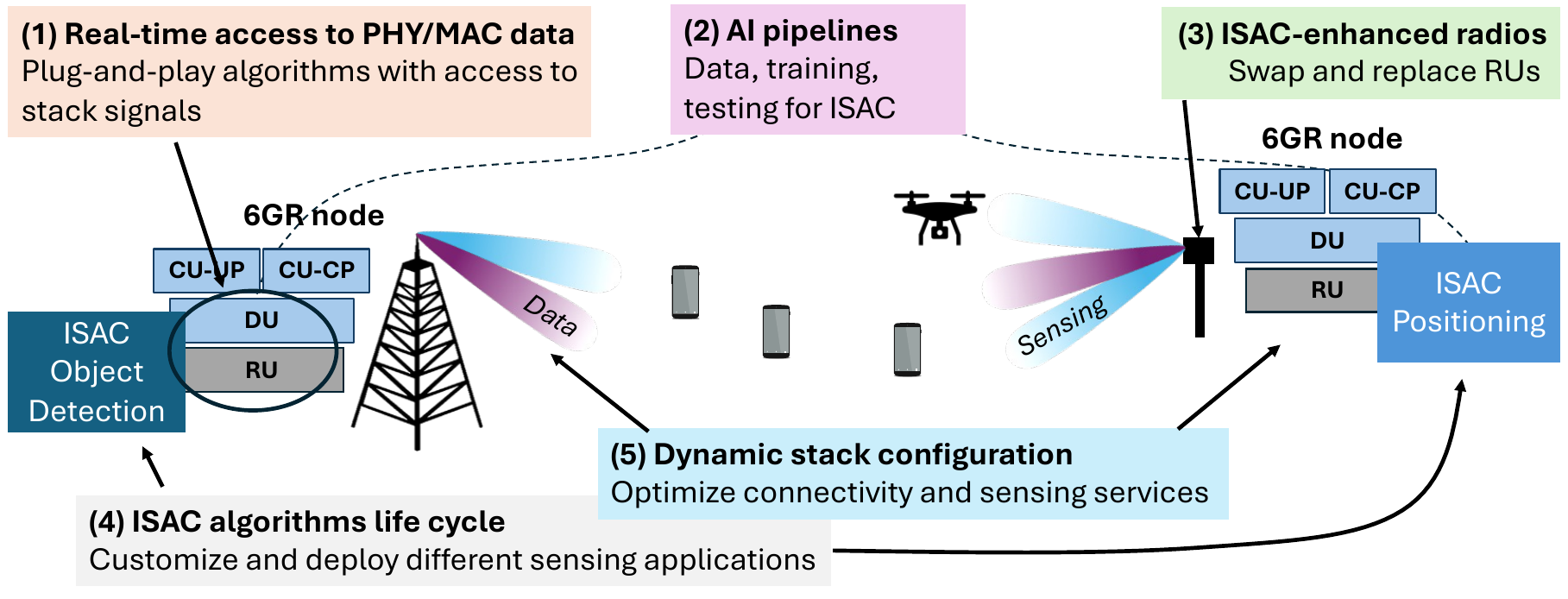}
\caption{System-level challenges toward enabling programmable \gls{isac} and inference within the 6GR architecture.}
% \vspace{-.3cm}
\label{fig:system_level_challenges}
\end{figure*}

The \gls{6g} of cellular networks is being designed in a radically different context compared to \gls{5g}. The pervasiveness of \gls{ai} brings new opportunities and challenges for both revenue and technical design. 
By deploying networks that support services beyond connectivity, operators aim at opening new revenue streams. Notable examples are sensing---with the network representing the physical world in a digital domain---and \gls{ai} inference---with complex models identifying structure and context in rich \gls{ran} data for optimization and value-added services.

The 3GPP is defining the next-generation \gls{ran} \rev{(6G Radio, or 6GR)} with sensing and \gls{ai} at the forefront.  
\gls{isac} allows multiplexing of communications and sensing on the same air interface, spectrum, and infrastructure. Physical-layer \gls{isac} has been widely researched~\cite{fatemisac2025,wei2023integrated}, and 3GPP is advancing study items for Release 20, due in 2028. 
Prior architectural work has explored sensing integration in cellular networks~\cite{gersing2024architecture,hamidi2024integrating,pan2025nextg}, but has largely focused on single-node topologies, waveform design, or core-network interfaces---leaving open questions on algorithm life-cycle management, real-time I/Q-level data access, and multi-node sensing coordination at the \gls{ran} level.

A programmable approach is essential: it allows operators and third parties to deploy, update, and customize sensing algorithms per site and application, enables \gls{ai}-driven sensing pipelines with closed-loop training workflows, and supports dynamic trade-offs between communication and sensing resources.

\textbf{System Challenges.} 
Figure~\ref{fig:system_level_challenges} summarizes five system-level challenges:
\emph{(1) Real-time PHY/MAC data access:} sensing algorithms require direct access to I/Q samples, \gls{csi}, and \gls{srs}, but current O-RAN interfaces like E2 lack the bandwidth and latency for user-plane data exposure at sensing timescales, especially when considering aggregating data over multiple base stations and users~\cite{d2022dapps}.
\emph{(2) AI pipelines:} AI-driven \gls{isac} requires end-to-end MLOps for data collection (with a focus on the data sources discussed above), training, validation, and deployment, not fully developed within open and standardized frameworks in the current \gls{ran} ecosystem.
\emph{(3) ISAC-enhanced radios:} different sensing topologies impose different RU requirements (full-duplex, tight synchronization, flexible beamforming); the architecture must support swappable, heterogeneous RUs.
\emph{(4) Algorithm life cycle:} sensing applications vary across sites and time; the architecture must support dynamic deployment and replacement of algorithms, as O-RAN does for xApps/rApps today.
\emph{(5) Dynamic stack configuration:} operators need to dynamically allocate resources between communication and sensing based on demand, policy, or regulation.

\textbf{Our Contribution.}
The Open \gls{ran} paradigm, with disaggregation, open interfaces, and programmable control loops, provides a natural foundation. The AI-RAN vision extends it along three axes: \emph{AI-for-RAN} (AI improving network functions), \emph{AI-on-RAN} (AI edge services on RAN infrastructure), and \emph{AI-and-RAN} (infrastructure sharing between AI and RAN workloads). \rev{Programmable \gls{isac} spans all three: sensing algorithms enhance RAN awareness (AI-for-RAN), deliver value-added services to third parties (AI-on-RAN), and share accelerated infrastructure with the communication stack (AI-and-RAN).} 

We argue that dApps---programmable applications, designed to be co-located with \gls{ran} functions, e.g., the DU, to provide real-time user-plane access---and a hierarchical control architecture are the key enablers for programmable \gls{isac} in both 5G NR and the upcoming 6GR. Concretely, we propose a system design in which sensing dApps access I/Q samples, \gls{csi}, and \gls{srs} data through a new E3 interface at the DU, running multiple plug-and-play sensing functions in parallel (e.g., monostatic, bistatic, ranging) alongside the communication stack on shared accelerated infrastructure. These edge-level dApps feed processed measurements to sensing xApps in the Near-RT \gls{ric} for multi-node fusion and collaborative inference, while an ISAC Orchestrator in the \gls{smo} manages the full algorithm life cycle---from centralized training and model cataloging to intent-driven, per-site deployment and continuous performance monitoring. This design builds on O-RAN open interfaces and established system components and \rev{workflows} and extends them to support programmable \gls{isac} flows.

We present this architecture and its components, and provide experimental results demonstrating the value of real-time observability for sensing accuracy and latency.

\begin{table*}[ht]
\centering
\caption{System-level challenges for programmable \gls{isac} and how the proposed architecture addresses them.}
\label{tab:challenges}
\begin{tabular}{p{0.18\textwidth} p{0.37\textwidth} p{0.37\textwidth}}
\toprule
\textbf{Challenge} & \textbf{Current Limitation} & \textbf{Proposed Solution} \\
\midrule
(1) PHY/MAC data access & E2 designed for control-plane telemetry; insufficient bandwidth/latency for I/Q, \gls{csi} streaming & dApps at \gls{du} with E3 interface for direct I/Q, \gls{csi}, \gls{srs} access \\
\midrule
(2) AI pipelines & No integrated MLOps for sensing model training, validation, deployment & Centralized \gls{isac} training pipeline with model catalog and automated deployment via \gls{smo} \\
\midrule
(3) ISAC-enhanced radios & Monolithic \gls{ru} designs; limited full-duplex or heterogeneous topology support & Open Fronthaul with swappable \gls{ru}s (FD-RU, Massive MIMO, FR-2/3) matched to sensing topology \\
\midrule
(4) Algorithm life cycle & Static deployment; no per-site customization or dynamic updates & \gls{isac} Orchestrator in \gls{smo} with dynamic algorithm/site matching and intent-driven deployment \\
\midrule
(5) Stack and infrastructure configuration & Fixed communication/sensing resource partitioning & Hierarchical control (dApps, xApps, rApps) for dynamic resource trade-offs \\
\bottomrule
\end{tabular}
\end{table*}

%---------------------------------------------%
    \section{O-RAN and ISAC Background} \label{sec:background}
%---------------------------------------------%

\textbf{O-RAN and Programmable Control.} The 3GPP and \gls{oran} 5G architecture disaggregates the \gls{gnb} into modular components---\gls{ru}, \gls{du}, and \gls{cu}---connected via open interfaces (Open Fronthaul, F1, E1). The \gls{du} handles real-time L1/L2 processing, the \gls{cu} manages higher-layer functions, and the \gls{ru} performs RF operations. Beyond disaggregation, O-RAN introduces programmable control loops: the Near-RT \gls{ric} hosts xApps for near-real-time optimization (10\,ms--1\,s), and the Non-RT \gls{ric} with rApps handles policy and orchestration at longer timescales. This programmable framework exposes data and control to third-party applications through open interfaces~\cite{d2022dapps}.

However, these interfaces were designed for communication control, not sensing. The E2 interface provides control-plane telemetry, not user-plane I/Q streams, preventing observability of key data for sensing.
% creating a fundamental barrier to integrating sensing as a \gls{ran}-level service.

\textbf{\gls{isac} Fundamentals and Standardization.}
\gls{isac} reuses cellular infrastructure for environment sensing by leveraging communication waveforms (OFDM, OTFS) or jointly designed signals. Monostatic sensing uses the same node for TX/RX but faces self-interference; bistatic systems use separate nodes thus requiring additional network resources; multistatic configurations combine multiple links for enhanced detection through spatial diversity~\cite{fatemisac2025}. 
On the standardization front, 3GPP Release~19 established use cases, requirements, and channel models for \gls{isac}~\cite{3gpp_22_137,3gpp_22_837}, covering monostatic and bistatic modes. Release~20, launched in June 2025, includes study items on \gls{isac} configuration, measurements, architecture, and sensing data exposure APIs~\cite{3gpp_rp_2340697}, but remains focused on waveform design, channel models, and core network integration, rather than the operational elements addressed by the architecture proposed in this paper. 

The concept of dApps, originally proposed in~\cite{d2022dapps}, has been developed through an architecture~\cite{lacava2025dapps} and two research reports within the O-RAN ALLIANCE \gls{ngrg}~\cite{oran-ngrg-rr2024-10,oran-ngrg-rr2025-05}.
They mention \gls{isac} as a use case, without however discussing workflows and enablers for \gls{isac} \gls{lcm}.
Concurrent work~\cite{baena2026native} proposes O-RAN extensions for monostatic sensing, including a sensing-specific E2 service model (E2SM-SENS) and Open Fronthaul metadata for waveform-echo association. Our work is complementary: we address the full \gls{isac} system architecture across multiple sensing topologies, including Open Fronthaul support for heterogeneous \glspl{ru}, hierarchical coordination from dApps through xApps/rApps to core \glspl{nf}, and---critically---\gls{isac} life-cycle management with centralized training, model cataloging, and intent-driven deployment.
Bridging these gaps---particularly for real-time sensing data access, algorithm life-cycle management, and collaborative sensing coordination---is the focus of this paper.

%---------------------------------------------%
\section{Programmable Sensing in Open AI-RAN} \label{sec:archi}
%---------------------------------------------%

\begin{figure}[t]
\centering
\includegraphics[width=\linewidth]{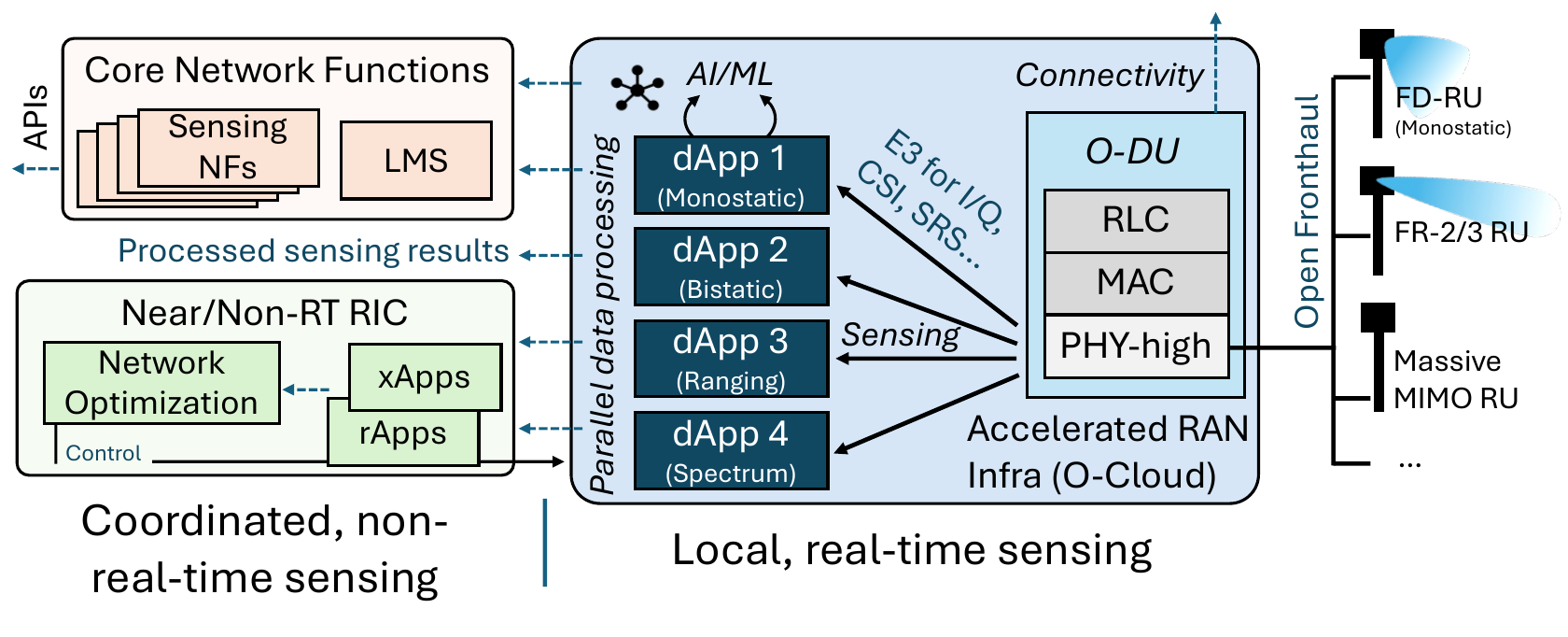}
\vspace{-.6cm}
\caption{Hierarchical sensing architecture. 
\rev{dApps provide local, real-time sensing at the \gls{du} via the E3 interface for I/Q, \gls{csi}, and \gls{srs} data. Multiple dApps run in parallel for different sensing topologies (monostatic, bistatic, ranging, spectrum). The Near/Non-RT \gls{ric} coordinates multi-node fusion and network optimization through sensing xApps and rApps. Processed results are exposed to core network sensing \gls{nf}s and external APIs.}
}
% \vspace{-.3cm}
\label{fig:system_sensing}
\end{figure}

We propose an enhancement to the \gls{oran} architecture, illustrated in Figure~\ref{fig:system_sensing}, based on four design principles: \emph{latency-optimized data flow} (routing data based on application timescale requirements), \emph{hierarchical processing} (lightweight edge algorithms with network-level fusion), \emph{parallel, asynchronous data handling} (parallel processing for multi-algorithm data processing), and \emph{\gls{ai} integration} (MLOps pipelines for training, versioning, and monitoring). Table~\ref{tab:challenges} summarizes how architectural components map to the challenges identified in Section~\ref{sec:intro}.

\subsection{Edge-Level Sensing with dApps}

In the current O-RAN framework, the data plane is opaque to external applications. Sensing algorithms operating at the physical layer require I/Q samples, \gls{csi}, and \gls{srs} at sub-millisecond timescales---data that the E2 interface was never designed to carry. Without direct user-plane access, real-time sensing is architecturally infeasible. dApps deployed at the \gls{du} level address this gap directly~\cite{d2022dapps,lacava2025dapps}.

For processing of lower layer user plane signals, dApps operate alongside the \gls{du} \gls{rlc}, \gls{mac}, and PHY-high functions, accessing I/Q samples, \gls{csi}, and \gls{srs} measurements through the E3 interface.
As Figure~\ref{fig:system_sensing} shows, in the dApp-driven \gls{isac} architecture
multiple dApps can run in parallel. Each dApp can implement a different sensing mode, or different algorithms for the same sensing mode \rev{focused} on different targets. This would allow a government-focused sensing algorithm to share data input with a commercial sensing solution, implemented in two different dApps. Examples are shown in Figure~\ref{fig:system_sensing}: a monostatic dApp with data from a full-duplex \gls{ru}, a bistatic dApp correlating signals from neighboring nodes, a ranging dApp estimating distances from time-of-arrival, or a spectrum sensing dApp for dynamic sharing. This plug-and-play model lets operators deploy, update, and replace sensing algorithms independently of the communication stack.

dApps perform edge-level real-time inference optimized for the local radio configuration. They can leverage co-located \gls{ai} accelerators for inference tasks such as \gls{cnn}-based target classification from I/Q inputs~\cite{lacava2025dapps}. Critically, sensing dApps can also feed back to the communication stack: scatterer information and channel conditions from sensing improve beam selection, channel estimation, and scheduling. This bidirectional coupling at the edge is a key advantage of the architecture.

The O-Cloud infrastructure underpinning the \gls{du} plays a central role in enabling this design. Sensing dApps and communication stack functions share the same accelerated hardware---\gls{gpu}s, NPUs, or FPGAs---provisioned through the O2 interface between the \gls{smo} and the O-Cloud platform. The O-Cloud must therefore schedule compute, memory, and I/O resources so that sensing inference does not starve the real-time communication pipeline, and vice versa. This is a concrete instance of Challenge~5 (dynamic stack configuration): as sensing workloads scale---e.g., multiple parallel dApps or increasingly complex \gls{ai}---the O-Cloud resource manager must dynamically partition accelerator capacity between sensing and connectivity, guided by policies pushed from the \gls{isac} Orchestrator described in Section~\ref{sec:lcm}.
\rev{Candidate mechanisms include container-level resource quotas, \gls{gpu} time-slicing or hardware partitioning (e.g., MIG), and priority-based scheduling that preempts sensing workloads when communication demand spikes.
Security of E3 data exposure is also critical: access control and data isolation mechanisms must prevent unauthorized exfiltration of I/Q samples by compromised or malicious dApps.}

\subsection{Network-Level Coordination}

Collaborative sensing demands data fusion across asynchronous, heterogeneous nodes. The data volumes involved require hierarchical processing: lightweight edge algorithms for local detection and network-level fusion for multi-node coordination. No standardized framework exists for this in O-RAN today. The Near-RT \gls{ric}, with its existing service framework for xApps, provides a natural platform to fill this role.

Sensing xApps at the Near-RT \gls{ric} coordinate multi-node operations, consuming processed measurements from dApps---compressed detection reports, estimated target parameters, feature vectors---rather than raw I/Q, keeping bandwidth manageable while enabling network-level fusion. This opens the opportunity to introduce advanced fusion algorithms handling measurement uncertainties, synchronization errors, and varying data quality across nodes to produce unified sensing outputs (tracked trajectories, environmental maps, event notifications).

xApps also manage collaborative resource allocation: coordinating sensing scheduling, beamforming, and spectrum use across nodes while respecting communication \gls{qos} requirements. The interaction between dApps and xApps uses the E2 interface extended with sensing-specific service models. At longer timescales, rApps in the Non-RT \gls{ric} set sensing policies, and sensing \gls{nf}s in the core expose results through standardized APIs.
\rev{Waveform configuration for sensing---e.g., \gls{srs} periodicity, bandwidth allocation, beam patterns---is managed by xApps for multi-node coordination or by rApps for policy-level decisions, while dApps consume and process the resulting signals.}

\subsection{\gls{isac} Life-Cycle Management} \label{sec:lcm}

\begin{figure*}[t]
\centering
\includegraphics[width=.7\linewidth]{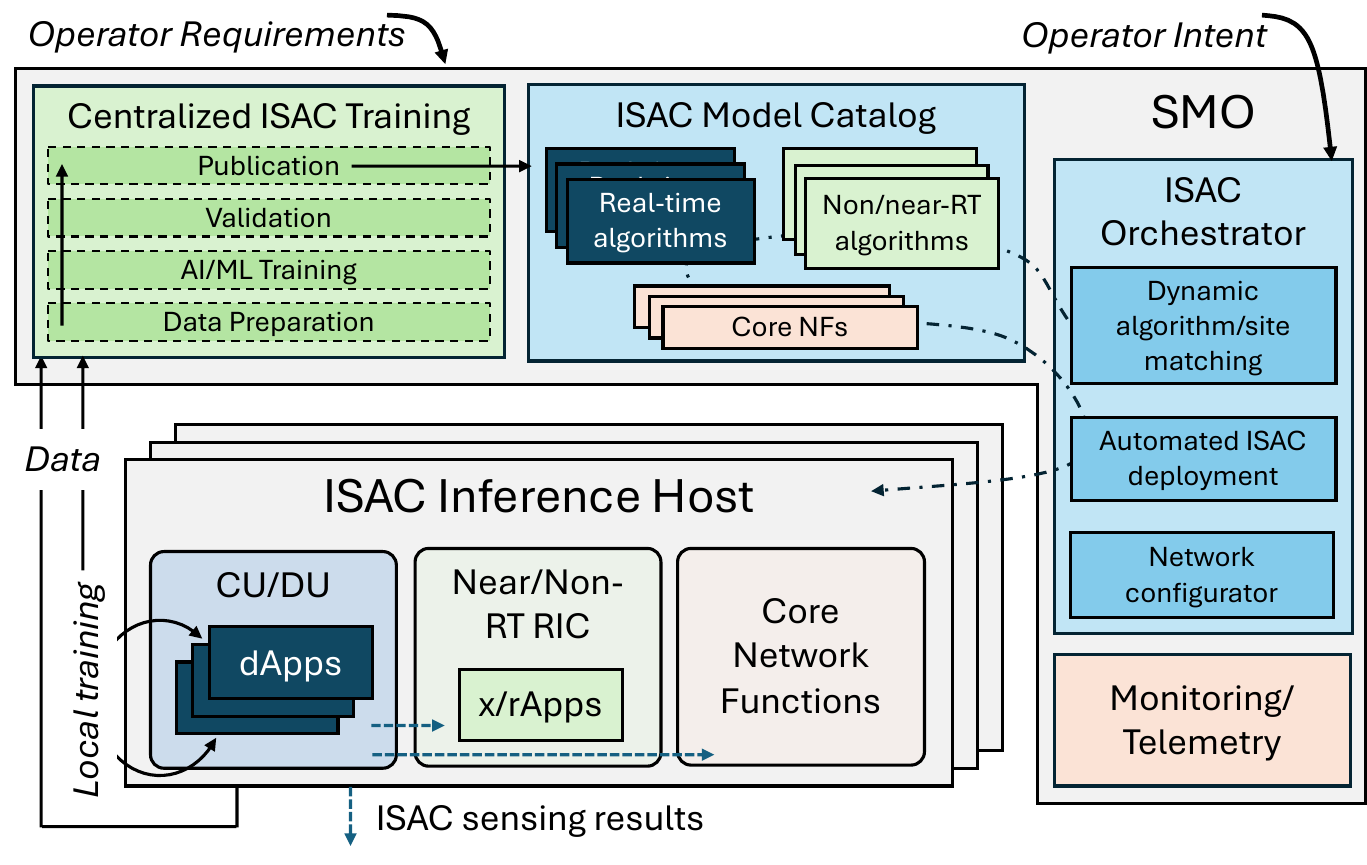}
\caption{\gls{isac} life-cycle management framework. 
\rev{Centralized training produces validated models published to the \gls{isac} Model Catalog. The \gls{isac} Orchestrator in the \gls{smo} performs intent-driven algorithm/site matching and automated deployment. Monitoring feeds enable continuous adaptation and retraining.}
}
% \vspace{-.3cm}
\label{fig:operations}
\end{figure*}

The algorithm life-cycle problem has key challenges related to sensing. Unlike communication control loops, sensing algorithms may require site-specific training data, specialized hardware acceleration, and continuous adaptation to changing physical environments. This is in addition to the possibility of generating different sensing routines from the same data, focusing on different customer requirements that the network operator may want to satisfy (e.g., as previously discussed, government vs. commercial sensing solutions). A sensing dApp for drone detection in a rural macro cell differs fundamentally from one for indoor positioning at an FR-2 small cell---yet both must be managed within the same architectural framework, and connect to the same, or similar, RAN infrastructure and implementation. Figure~\ref{fig:operations} illustrates the proposed \gls{lcm} framework that addresses this challenge.

\textbf{Centralized training and model catalog.} A training pipeline within the \gls{smo} handles data preparation, \gls{ai} training, validation, and publication to an \gls{isac} Model Catalog. The catalog organizes validated algorithms by function (dApp vs. xApp/rApp), sensing topology, target application, and deployment requirements. Local training at individual sites complements centralized training, enabling adaptation to site-specific conditions using data collected by dApps during operation.

\textbf{\gls{isac} Orchestrator.} Residing in the \gls{smo}, the Orchestrator translates operator intent---e.g., ``enable drone detection across rural macro sites''---into deployment actions. It performs dynamic algorithm/site matching based on each site's radio configuration, compute resources, and sensing requirements, then pushes selected dApps, xApps, and core \gls{nf}s through O-RAN orchestration interfaces (O1, O2, A1). Deployment through O2 is particularly important: the Orchestrator must verify that the target O-Cloud instance has sufficient accelerator capacity (e.g., \gls{gpu} cores, memory) to host the sensing dApp alongside existing communication workloads before instantiating it, and may trigger O-Cloud scaling or workload migration when resources are insufficient.

\textbf{Monitoring and adaptation.} The Orchestrator continuously monitors deployed algorithms via telemetry, tracking detection probability, false alarm rate, localization accuracy, and latency. When performance degrades, it triggers retraining, model replacement, or reconfiguration. The ISAC Orchestrator coordinates with the SMO and RICs to configure the protocol stack.

This framework transforms \gls{isac} from a static, vendor-defined capability into a dynamic, operator-managed service.

%---------------------------------------------%
\section{Numerical Results} \label{sec:results}
%---------------------------------------------%

To motivate the proposed architecture---particularly the need for real-time I/Q exposure through the E3 interface---we present two results that map to the monostatic and ranging dApps in Figure~\ref{fig:system_sensing}. The first uses a \gls{crlb} analysis to quantify the overhead associated with data that needs to be exposed and processed for sensing; the second reports experimental results from an \gls{oai}-based \gls{5g} testbed demonstrating the value of dApp-level access to full \glspl{cir} data. \rev{The first result maps to a monostatic full-duplex topology (dApp~1); the second to network-assisted uplink ranging (dApp~3)---together illustrating how dApps serve fundamentally different sensing modes through the same E3 interface.}

\subsection{Monostatic Sensing: CRLB vs.\ Data Movement Overhead}

We quantify the fundamental trade-off facing a full-duplex monostatic dApp (dApp~1 in Figure~\ref{fig:system_sensing}): how much bandwidth (e.g., on the E3 interface) must be allocated to I/Q transfer to achieve a given sensing accuracy?
We model an \gls{ofdm} waveform at $f_c = 3.6$\,GHz with 30\,kHz subcarrier spacing and 7\% cyclic prefix overhead. After matched filtering, the received signal on the $k$-th subcarrier of the $m$-th symbol is a 2-D complex sinusoid whose frequencies encode the target delay and Doppler~\cite{braun2014ofdm}. The \gls{crlb} for delay and Doppler estimation then provides the minimum achievable range and velocity \gls{rmse} for any unbiased estimator, as a function of the per-resource-element SNR, the number of subcarriers $N$, and the number of symbols $M$.
The target is a small drone at 500\,m range, 25\,m/s radial velocity, and $-20$\,dBsm \gls{rcs}, with 43\,dBm transmit power. Bandwidth (and thus number of subcarriers) and slot duration (and thus number of symbols) are swept jointly from 10 to 100\,MHz and 0.25 to 1.0\,ms, respectively.

% --- FIGURE: CRLB vs Overhead (dual y-axis, single plot) ---
% Data generated from fd_crlbVsOvhead.m (MATLAB) / equivalent Python script.
% The x-axis "sensing overhead" is the E3 I/Q data rate: M*N*b*8 / T_slot [Mbps].
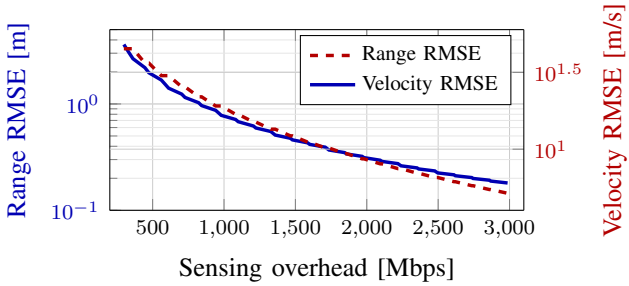
\begin{figure}[t]
\centering
\begin{tikzpicture}
\pgfplotsset{
    every axis/.style={
        width=0.8\columnwidth,
        height=0.45\columnwidth,
        xmin=200, xmax=3100,
        xlabel={Sensing overhead {[Mbps]}},
        grid=both,
        grid style={line width=0.2pt, draw=gray!20},
        major grid style={line width=0.3pt, draw=gray!40},
        tick label style={font=\scriptsize},
        label style={font=\small},
        legend style={font=\scriptsize, at={(0.97,0.97)}, anchor=north east, cells={anchor=west}},
    },
}
% Left y-axis: Range RMSE
\begin{semilogyaxis}[
    axis y line*=left,
    ylabel={Range RMSE {[m]}},
    ylabel style={color=blue!70!black},
    yticklabel style={color=blue!70!black},
    ymin=0.1, ymax=5,
]
\addplot[color=blue!70!black, solid, line width=1.3pt, mark=none]
    table[col sep=comma, x=overhead_mbps, y=rmse_range_m]{DATA/crlb_data.csv};
\label{pgf:range}
\end{semilogyaxis}
% Right y-axis: Velocity RMSE
\begin{semilogyaxis}[
    axis y line*=right,
    axis x line=none,
    ylabel={Velocity RMSE {[m/s]}},
    ylabel style={color=red!70!black},
    yticklabel style={color=red!70!black},
    ymin=4, ymax=60,
]
\addplot[color=red!70!black, dashed, line width=1.3pt, mark=none]
    table[col sep=comma, x=overhead_mbps, y=rmse_vel_ms]{DATA/crlb_data.csv};
\label{pgf:vel}
\addlegendimage{/pgfplots/refstyle=pgf:range}\addlegendentry{Range RMSE}
\addlegendentry{Velocity RMSE}
\end{semilogyaxis}
\end{tikzpicture}
\caption{\small\gls{crlb}-based sensing accuracy vs.\ data movement overhead for full-duplex monostatic sensing. A drone target at 500\,m range, 25\,m/s velocity, $-20$\,dBsm \gls{rcs}. Bandwidth (10--100\,MHz) and slot duration (0.25--1.0\,ms) increase jointly and are represented by the sensing overhead rate in the x axis.}
\label{fig:crlb_overhead}
\end{figure}

Figure~\ref{fig:crlb_overhead} shows the \gls{crlb}-derived \gls{rmse} for range and velocity as a function of the sensing data-rate overhead, defined as $\mathcal{O} = M N b \cdot 8 / T_{\text{slot}}$\,[Mbps], where $b=4$ bytes per I/Q sample (16-bit I + 16-bit Q). This overhead represents the throughput required to transfer raw I/Q samples from the DU PHY to the co-located sensing dApp through the E3 interface.
Both curves decrease monotonically: at 300\,Mbps the range \gls{rmse} is approximately 3.6\,m, improving to below 0.2\,m at 3\,Gbps; velocity \gls{rmse} drops from 45\,m/s to about 5\,m/s over the same range. These results directly quantify that it is not conceivable to offload monostatic sensing to components that are external to a RAN site, and motivates the adoption of a local interface such as E3, especially as multiple RAN nodes are considered. Further, the sensing overhead competes with communication data plane resources, requiring the hierarchical control loops to balance sensing and connectivity in real time (Challenge~5 in Table~\ref{tab:challenges}).

\subsection{Ranging dApp: Experimental Validation}

% --- FIGURE: Ranging CDF (experimental data from OAI-based 5G testbed) ---
% Data extracted from range_dapp.fig (MATLAB) -> DATA/ranging_cdf_data.csv
\begin{figure}[t]
\centering
\begin{tikzpicture}
\begin{axis}[
    width=0.92\columnwidth,
    height=0.5\columnwidth,
    xlabel={Range estimation error {[m]}},
    ylabel={Empirical CDF},
    xmin=0, xmax=2,
    ymin=0, ymax=1.05,
    grid=both,
    grid style={line width=0.2pt, draw=gray!30},
    major grid style={line width=0.4pt, draw=gray!50},
    tick label style={font=\scriptsize},
    label style={font=\small},
    legend style={font=\tiny, at={(0.97,0.4)}, anchor=east, cells={anchor=west}},
    legend cell align={left},
    line width=1.1pt,
]
% Individual Peak report, M=20 (blue, dashed)
\addplot[color=blue!80!black, dashed, mark=none]
    table[col sep=comma, x=range_error_m, y=peak_m20]{DATA/ranging_cdf_data.csv};
\addlegendentry{Individual Peak report (LMF), $M\!=\!20$}
% Individual Peak report, M=60 (blue, solid)
\addplot[color=blue!80!black, solid, mark=none]
    table[col sep=comma, x=range_error_m, y=peak_m60]{DATA/ranging_cdf_data.csv};
\addlegendentry{Individual Peak report (LMF), $M\!=\!60$}
% Subspace method + dApp, M=20 (red, dashed)
\addplot[color=red!80!black, dashed, mark=none]
    table[col sep=comma, x=range_error_m, y=subspace_m20]{DATA/ranging_cdf_data.csv};
\addlegendentry{Subspace method + dApp, $M\!=\!20$}
% Subspace method + dApp, M=60 (red, solid)
\addplot[color=red!80!black, solid, mark=none]
    table[col sep=comma, x=range_error_m, y=subspace_m60]{DATA/ranging_cdf_data.csv};
\addlegendentry{Subspace method + dApp, $M\!=\!60$}
\end{axis}
\end{tikzpicture}
\caption{\small CDF of range estimation error from an \gls{oai}-based \gls{5g} testbed. A subspace method running as a dApp on full \gls{cir} data accessed via the E3 interface substantially outperforms the scalar peak-detection pipeline used in current \gls{5g} \gls{nr} positioning, especially with fewer observations ($M=20$)~\cite{gangula2025rtt}.}
\label{fig:ranging_cdf}
\end{figure}
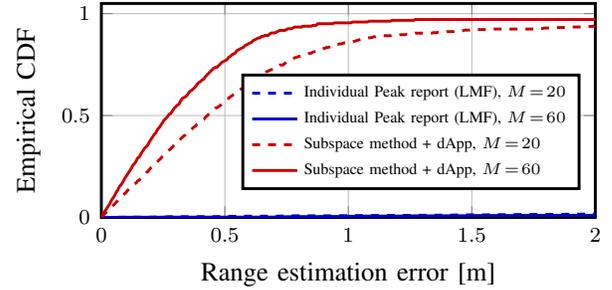

We validate the dApp concept experimentally using a ranging dApp (dApp~3 in Figure~\ref{fig:system_sensing}) implemented on an \gls{oai}-based \gls{5g} testbed~\cite{gangula2025rtt}. \rev{The testbed operates at 3.6\,GHz with 40\,MHz bandwidth in an indoor environment with rich multipath, using a single antenna at the \gls{gnb}.} In current \gls{5g} \gls{nr}, the \gls{gnb} estimates the uplink \gls{cir} from the \gls{srs} transmitted by the \gls{ue}. The physical layer applies peak detection on each individual \gls{cir} estimate to extract a scalar timing measurement (e.g., UL-RTOA for UL-TDoA, or gNB Rx-Tx time difference for multi-RTT), which is reported to the Location Management Function (LMF) in the core network. The full complex-valued \gls{cir}---containing amplitude, phase, and multipath structure---is discarded at the PHY boundary.

A ranging dApp co-located with the \gls{du} instead retains $M$ successive \gls{cir} observations through the E3 interface and applies a subspace-based estimation method (e.g., MUSIC) jointly across snapshots. This enables detection of the line-of-sight delay even when the corresponding path is indistinguishable from noise in any single CIR observation.
% This enables super-resolution of the line-of-sight delay from nearby multipath, which peak detection on individual observations cannot achieve.

Figure~\ref{fig:ranging_cdf} shows the empirical CDF of range estimation error. Two approaches are compared: scalar peak detection (current \gls{5g} pipeline, where $M$ individual reports are sent to the LMF) and the subspace method running as a dApp. With $M=20$ observations, the subspace dApp achieves sub-meter accuracy for over 90\% of estimates, while the scalar peak-detection method completely fails to estimate the range. Increasing to $M=60$ further tightens the subspace distribution. Critically, the peak-detection approach gains comparatively little from additional observations---it cannot recover the signal \rev{subspace} as the CIR with multipath components is discarded at the PHY boundary. These experimental results validate the core premise of the dApp architecture: real-time access to full physical-layer data through E3 enables sensing algorithms that are fundamentally impossible with scalar \gls{kpm} reporting via E2.

%---------------------------------------------%
\section{Conclusion and Open Challenges} \label{sec:concl}
%---------------------------------------------%

This paper presented an architectural framework for programmable inference and \gls{isac} at the 6GR edge. We identified five system-level challenges that current O-RAN and 3GPP frameworks do not address, and proposed a hierarchical sensing architecture where dApps provide real-time, I/Q-level sensing at the \gls{du} through an E3 interface, while xApps and rApps coordinate multi-node fusion and policy. An \gls{isac} life-cycle management framework in the \gls{smo} supports centralized training, model cataloging, and intent-driven deployment. Numerical and experimental results demonstrated that I/Q-level access and subspace processing enabled by dApps yield substantially better sensing accuracy than aggregated telemetry or scalar reporting, validating dApps as the key architectural enabler for transforming sensing into a dynamic, programmable 6GR service.

Several open challenges remain. On the \emph{technical} side, synchronization across distributed sensing nodes is critical for coherent multi-node processing, and the E3 interface specification---data formats, access patterns, isolation from the communication pipeline---requires further design. \emph{Standardization} efforts must bridge 3GPP sensing service definitions with O-RAN programmability (dApps, E3), potentially through joint study items and standardized sensing data representations for multi-vendor interoperability. \emph{AI-RAN coexistence} poses resource-allocation challenges: sensing dApps and \gls{ai} inference models compete with the communication stack for \gls{gpu} cycles, memory, and I/O on shared O-Cloud infrastructure, requiring mechanisms to prevent \gls{qos} degradation. 
\rev{Finally, \emph{privacy and security} remain open concerns: \gls{isac} can inadvertently collect personal information through passive sensing, and exposing raw I/Q data through the E3 interface creates a new attack surface. Compliance with privacy regulations, privacy-preserving algorithms, access control for third-party dApps, and data isolation at each layer of the sensing pipeline must be addressed~\cite{wei2023integrated}.}

\bibliographystyle{IEEEtran}
\bibliography{References}

% AUTHOR BIOS (max 150 words each)
% \begin{IEEEbiography}[{\includegraphics[width=1in,height=1.25in,clip,keepaspectratio]{photo.jpg}}]{Author Name}
% Bio text here (max 150 words).
% \end{IEEEbiography}

\vspace{-40pt}
\begin{IEEEbiographynophoto}
{Michele Polese} is a Research Assistant Professor at Northeastern University. He received his Ph.D. from the University of Padova in 2020.

\vspace{5pt}
\noindent
\textbf{Rajeev Gangula} is a Research Assistant Professor at Northeastern University. He received his Ph.D. from Telecom ParisTech (Eurecom), France, in 2015.

\vspace{5pt}
\noindent
\textbf{Tommaso Melodia} is the William Lincoln Smith Chair Professor at Northeastern University and Director of the Institute for Intelligent Networked Systems.
\end{IEEEbiographynophoto}

\end{document}